\documentclass[preprint,final,5p,times,twocolumn]{elsarticle}

\usepackage{hyperref}
\hypersetup{colorlinks,linkcolor=blue,anchorcolor=blue,citecolor=blue,urlcolor=blue}
\usepackage{graphicx}
\usepackage{epstopdf}
\usepackage{dcolumn}
\usepackage{textcomp}
\usepackage{gensymb}
\usepackage{booktabs}
\usepackage{amsmath}
\usepackage{amssymb}
\usepackage{color}
\usepackage{url}
\usepackage{siunitx}
\usepackage{subfigure}
\usepackage{longtable}
\usepackage{array}
\usepackage{booktabs}
\usepackage{verbatim}
\usepackage{threeparttable}

\emergencystretch=\maxdimen
\usepackage{dcolumn}
\newcolumntype{x}{D{.}{.}{6.6}}
\newcolumntype{y}{D{.}{.}{4.5}}
\newcolumntype{z}{D{.}{.}{5.7}}
\newcolumntype{f}{D{.}{.}{7.9}}
\newcolumntype{e}{D{.}{.}{5.6}}

\journal{Physics Letters B}

\begin{document}

\begin{frontmatter}


\title{Electromagnetic moments of scandium isotopes and \mbox{$N=28$} isotones in the distinctive $0f_{7/2}$ orbit}

\author[PKU]{S.~W.~Bai}
\author[KULEUVEN,LIVEPOOL]{\'{A}.~Koszor\'{u}s\fnref{fn1}}
\author[TRIUMF,PKU]{B.~S.~Hu}
\author[PKU]{X.F.~Yang\corref{cor1}}\ead{xiaofei.yang@pku.edu.cn}
\author[TUM]{J.~Billowes}
\author[TUM]{C.~L.~Binnersley}
\author[TUM]{M.~L.~Bissell}
\author[MPI]{K.~Blaum}
\author[TUM]{P.~Campbell}
\author[LIVEPOOL]{\mbox{B.~Cheal}}
\author[KULEUVEN]{T.~E.~Cocolios}
\author[JYL]{R.~P.~de~Groote\fnref{fn2}}
\author[LIVEPOOL]{C.~S.~Devlin}
\author[TUM,TUM2]{\mbox{K.T.~Flanagan}}
\author[MIT,CERN]{\mbox{R.~F.~Garcia Ruiz}}
\author[CERN]{H.~Heylen}
\author[TRIUMF,TRIUMF2]{J.~D.~Holt}
\author[KULEUVEN]{A.~Kanellakopoulos}
\author[TUD]{J.~Kr\"{a}mer}
\author[CERN]{V.~Lagaki}
\author[TUD]{B.~Maa\ss{}}
\author[CERN]{S.~Malbrunot-Ettenauer}
\author[TRIUMF]{T.~Miyagi}
\author[MPI, MAINZ]{R.~Neugart}
\author[KULEUVEN,CERN]{G.~Neyens}
\author[TUD]{W.~N\"{o}rtersh\"{a}user}
\author[MPI,CERN,IPN]{L.~V.~Rodr\'{i}guez}
\author[TUD]{F.~Sommer}
\author[TUM]{A.~R.~Vernon}
\author[PKU]{S.~J.~Wang}
\author[HZ]{X.~B.~Wang}
\author[CERN2]{S.~G.~Wilkins}
\author[KULEUVEN]{Z.~Y.~Xu\fnref{fn3}}
\author[SYSU]{C.~X.~Yuan}

\cortext[cor1]{Corresponding author}
\fntext[fn1]{Present address: Experimental Physics Department, CERN, CH-1211 Geneva 23, Switzerland}
\fntext[fn2]{Present address: KU Leuven, Instituut voor Kern- en Stralingsfysica, B-3001 Leuven, Belgium}
\fntext[fn3]{Present address: Department of Physics and Astronomy, University of Tennessee, 37996 Knoxville, TN, USA}

\address[PKU]{School of Physics and State Key Laboratory of Nuclear Physics and Technology, Peking University, Beijing 100871, China}

\address[KULEUVEN]{KU Leuven, Instituut voor Kern- en Stralingsfysica, B-3001 Leuven, Belgium}

\address[LIVEPOOL]{Oliver Lodge Laboratory, Oxford Street, University of Liverpool, Liverpool, L69 7ZE, United Kingdom}

\address[TRIUMF]{TRIUMF, 4004 Wesbrook Mall, Vancouver, BC V6T 2A3, Canada}

\address[TUM]{School of Physics and Astronomy, The University of Manchester, Manchester M13 9PL, United Kingdom}

\address[MPI]{Max-Planck-Institut f\"{u}r Kernphysik, D-69117 Heidelberg, Germany}

\address[JYL]{Department of Physics, University of Jyv\"askyl\"a, PB 35(YFL) FIN-40351 Jyv\"askyl\"a, Finland.}

\address[TUM2]{Photon Science Institute Alan Turing Building, University of Manchester, Manchester M13 9PY,United Kingdom}

\address[MIT]{Massachusetts Institute of Technology, Cambridge, MA, USA}

\address[CERN]{Experimental Physics Department, CERN, CH-1211 Geneva 23, Switzerland}

\address[TRIUMF2]{Department of Physics, McGill University, 3600 Rue University, Montr\'eal, QC H3A 2T8, Canada}

\address[TUD]{Institut f\"{u}r Kernphysik, TU Darmstadt, D-64289 Darmstadt, Germany}

\address[MAINZ]{Institut f\"{u}r Kernchemie, Universit\"{a}t Mainz, D-55128 Mainz, Germany}

\address[IPN]{Institute de Physique Nucl\'eaire, CNRS-IN2P3, Universit\'e Paris-Sud,Universit\'e Paris-Saclay, 91406 Orsay, France}

\address[HZ]{School of Science, Huzhou University, Huzhou 313000, China}

\address[CERN2]{Engineering Department, CERN, CH-1211 Geneva 23, Switzerland}

\address[SYSU]{Sino-French Institute of Nuclear Engineering and Technology, Sun Yat-Sen University, Zhuhai, 519082, Guangdong, China}

\begin{abstract}
The electric quadrupole moment of $^{49}$Sc was measured by collinear laser spectroscopy at CERN-ISOLDE to be $Q_{\rm s}=-0.159(8)$~$e$b, and a nearly tenfold improvement in precision was reached for the electromagnetic moments of $^{47,49}$Sc. The single-particle behavior and nucleon-nucleon correlations are investigated with the electromagnetic moments of $Z=21$ isotopes and $N=28$ isotones as valence neutrons and protons fill the distinctive $0f_{7/2}$ orbit, respectively, located between magic numbers, 20 and 28. The experimental data are interpreted with shell-model calculations using an effective interaction, and \textit{ab-initio} valence-space in-medium similarity renormalization group calculations based on chiral interactions. These results highlight the sensitivity of nuclear electromagnetic moments to different types of nucleon-nucleon correlations, and establish an important benchmark for further developments of theoretical calculations.
\end{abstract}

\begin{keyword}
Collinear laser spectroscopy \sep Electromagnetic moments \sep Nucleon-nucleon correlation \sep Ab-initio calculation
\end{keyword}

\end{frontmatter}



\section{Introduction}
\label{sec1}

Since it was established by Mayer and Jensen~\cite{Mayer1949,Jensen1949}, the nuclear shell model~(SM) and the concept of magic numbers have played an essential role in our understanding of the structure of the nuclear quantum many-body system~\cite{Sorlin}. The independent-particle SM assumes non-interacting valence nucleons outside a spherical core, and can reasonably describe the properties of near-magic nuclei~\cite{Heyde1994,Neyens_2003}, such as their ground-state spins and electromagnetic moments. Deviations of the observed properties from this model are attributed to the residual nucleon-nucleon~(NN) interaction between the valence nucleons, and to the interaction of valence nucleon(s) with the core. Moments of nuclei with clear single-particle orbit configuration, measured with sufficient precision for a long range of isotopes, are sensitive probes of different aspects of the residual interaction, which can be included in large-scale SM calculations. While effective interactions used to be determined empirically for specific model spaces~\cite{TakaRevModPhys2020}, in the recent years, it has become possible to deduce more realistic interactions rooted in QCD, including two and three-body forces, through chiral effective field theory ($\chi$EFT)~\cite{Machleidt2011}.

More quantitatively, within the independent-particle SM, the magnetic moment for a single particle~(SP) occupying a SM orbit, the so-called \lq Schmidt moment', depends only on the angular momentum $j$ and the free nucleon magnetic moments~\cite{Neyens_2003}. It should thus remain constant as odd nucleons fill an orbit~\cite{Heyde1994,Neyens_2003}. Deviations from the Schmidt moments may be broadly attributed to two possible causes: configuration mixing of the wavefunctions and corrections of meson-exchange currents~(MEC) to the two-body magnetic-moment operator~\cite{Arima1954,Castel1990}. The quadrupole moment, on the other hand, is a good indicator of collective effects of the nucleus~\cite{Castel1990,Yoshina1972}. The SP quadrupole moment of a nucleon depends on the angular momentum $j$ and the mean-square charge radius of the orbit by the unpaired valence nucleons. As nucleons are added to an orbit, the seniority scheme of the independent-particle SM predicts quadrupole moments to follow a linear increase with the number of valence particles in the SM orbit, crossing zero at half filling~\cite{Heyde1994}. Some experimental linear trends have been observed in the Pb~(Cd) isotopes as neutrons fill the $\nu i_{13/2}$ ($\nu h_{11/2}$) orbit~\cite{Neyens_2003,Cd-moment} and in the $N=82$~($N=126$) isotones as protons fill the $\pi g_{7/2}$ ($\pi h_{9/2}$) orbit~\cite{simon-2021,Ferrer2017}. However, as these orbits are closely embedded among others in the shell, the scattering of the nucleons among several orbits may result in the zero-crossing of the linear trend away from half filling. A rather unexpected, and not yet explained deviation from such linear trend, was recently observed in the Sn isotopes~\cite{commphys-Sn}. \\
The $0f_{7/2}$ orbit, located between magic numbers 20 and 28, forms a unique example in the nuclear chart where a single orbit is well isolated from its neighbors. One can expect that electromagnetic moments of isotopes with valence protons and neutrons in the $0f_{7/2}$ orbit, e.g. the $N=28$ isotones and $Z=21$ isotopes, respectively, would be excellent probes to experimentally verify the SP nature and correlations. In addition, the moments of $N=28$ magic isotones would potentially offer an ideal platform to explore the seniority properties of the independent-particle SM~\cite{Heyde1994}, and the influence of $E2$ and $M1$ correlations. The electromagnetic moments of isotopes with such a simple configuration are also desired to validate the recent progress of \textit{ab-initio} many-body methods and microscopic interactions derived from $\chi$EFT. Those have been continuously improved to interpret nuclear masses and radii~\cite{Wienholtz2013,GarciaRuiz2016,Kradii} but so far have only scarcely been applied to the magnetic and quadrupole moments, another two basic properties of the atomic nucleus~\cite{Ca-moment2015, 37Ca-moment, Al-moment}.

This letter presents a precise measurement of the electromagnetic moments of the unstable nuclei $^{47,49}$Sc, yielding a first quadrupole moment for $^{49}$Sc. This provides key data to the systematics of nuclear moments associated with the $0f_{7/2}$ orbit and facilitates the investigation of the SP behavior and NN correlations. The experimental data are compared with shell-model calculations and valence-space in-medium similarity renormalization group~(VS-IMSRG) calculations~\cite{IMSRG-Tsukiyama2012,IMSRG-Stroberg2017} based on $\chi$EFT interactions~\cite{NNLOsat,NNLOgo}.

\section{Experimental method}
\label{sec2}

Because of conflicting data on magnetic moments from different measurement methods~\cite{Avgoulea_2011,47Sc}, as will be discussed further in \mbox{Sec.~\ref{sec3}}, two collinear laser spectroscopy~(CLS) setups, COLLAPS and CRIS, are adopted for this study. This allows the moments of Sc isotopes to be determined unambiguously from both atomic and ionic hyperfine structure~(hfs). Details on both setups can be found in Refs~\cite{Kradii,Neugart2017,tassos-Ge}. In brief, the Sc isotopes were produced by impinging 1.4-GeV protons onto a Ta-foil target at ISOLDE-CERN, and resonantly ionized with RILIS~\cite{RILIS}. The ions were accelerated up to 40~keV, mass separated, and cooled for 100~ms (or 10~ms) in a linear Paul trap~\cite{ISCOOL}. The Sc ions were released as bunches of $\sim$5-$\mu$s temporal length and sent to either the COLLAPS or CRIS set-up. At COLLAPS, the ion bunch was collinearly overlapped with a frequency-doubled continuous wave Ti:Sapphire laser at 364.3~nm to match the Doppler-shifted \mbox{$3d4s$ $^{3\!}D_{1}$} $\to$ \mbox{$3d4p$ $^{3\!}F_{2}$} ionic transition. The laser frequency was stabilized by a wavemeter, which was calibrated in real time by a diode laser locked to one hyperfine component of the $^{87}$Rb atom. The ion velocity was tuned to probe the resonant excitation of the transition. Four photomultiplier tubes were used to record the fluorescence photons emitted from the laser-excited Sc ions as a function of the scanning voltage to obtain the hfs spectrum. At CRIS, the ion bunches (100~Hz) were neutralized using a potassium vapor and then overlapped in time and space with two 100-Hz pulsed lasers at 246.8~nm and 532~nm, respectively. The first frequency-tripled narrow-band laser was used to resonantly excite the atoms via the \mbox{$3d4s^2$ $^{2\!}D_{3/2}$} $\to$ \mbox{$3d4s5p$ $^{2\!}P^0_{1/2}$} transition, and the subsequent frequency-doubled Nd:YAG laser to ionize them. The resonantly ionized ions were deflected from the beam, and recorded by an ion detector as a function of the laser frequency detuning to obtain the hfs spectrum~\cite{adamthesis}.

\begin{figure}[t!]
	\includegraphics[width=0.5\textwidth]{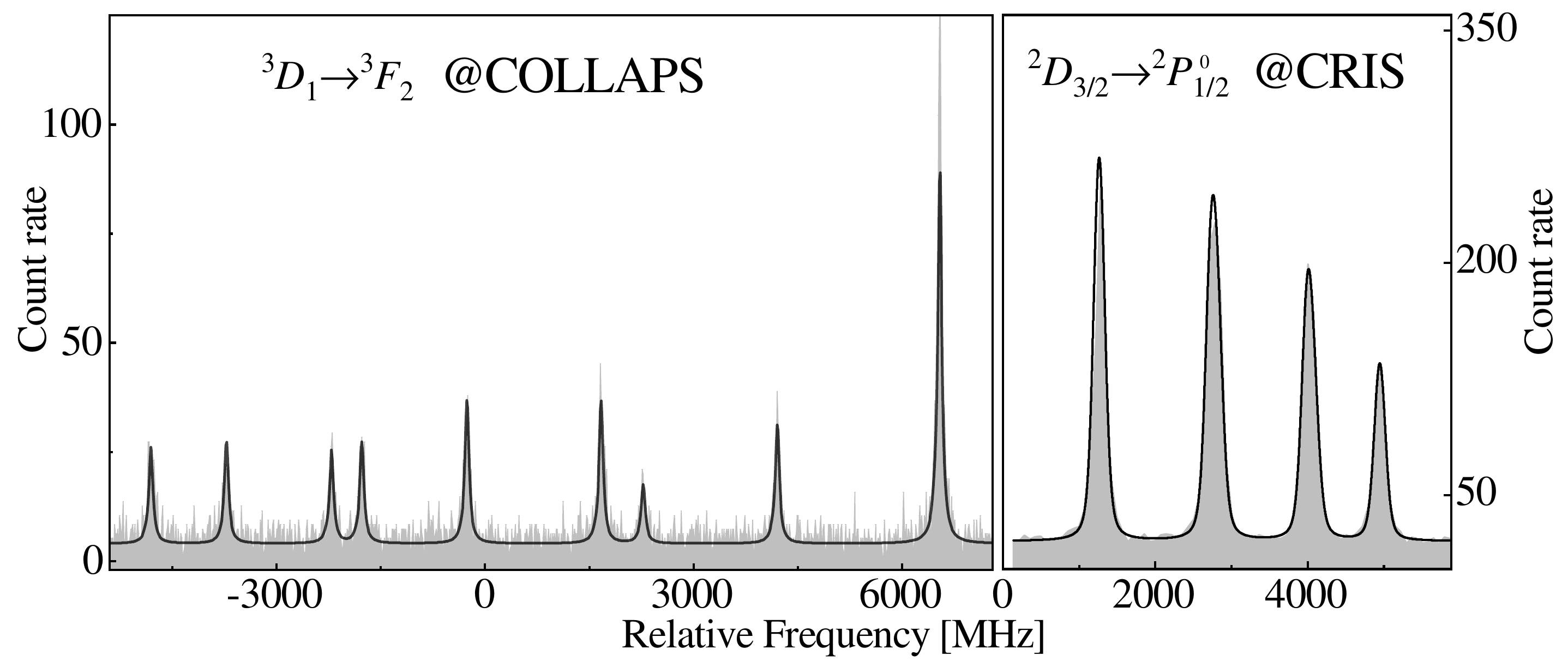}
	\caption{\label{fig:fig1}\footnotesize{Hfs spectra of $^{49}$Sc measured with the COLLAPS and CRIS methods on respectively ionic and atomic Sc beam. Data are fitted with a Voigt line profile using SATLAS~\cite{SATLAS}.}}
\end{figure}

\section{Experimental results}
\label{sec3}

Example hfs spectra of $^{49}$Sc measured with both methods are shown in Fig.~\ref{fig:fig1}, and analyzed using the $\chi^{2}$-minimization approach in SATLAS~\cite{SATLAS}. The extracted magnetic and quadrupole hfs parameters ($A$ and $B$) are summarised in Tab.~\ref{tab:table10} and in good agreement with the literature values~\cite{Avgoulea_2011,45Sc-AB,46Sc-AB}. The magnetic moments ($\mu$) of $^{46,47,49}$Sc were extracted from two ionic and one atomic $A$ parameter, using \mbox{$\mu= \mu_{\rm{45}}IA/(I_{\rm{45}}A_{\rm{45}}$)} with the re-evaluated $\mu$($^{45}$Sc)~\cite{45Sc-CPL}. The final $\mu$ measured with COLLAPS are calculated as the weighted average of the two sets of magnetic moments taking into account the correlation between \mbox{$A(^{3}D_{1})$ and $A(^{3}F_{2})$}. As presented in Tab.~\ref{tab:table11}, the magnetic moments measured with the two CLS methods are in excellent agreement with each other. The quadrupole moments ($Q_s$) are obtained from the larger $B(^{3}F_{2})$ parameters using \mbox{$Q_s= Q_{\rm s,45}B/B_{\rm{45}}$} with the most recent recommended value of $Q_{\rm s}$($^{45}$Sc)~\cite{Q-45Sc}.

\begin{table}[!htb]
	\begin{threeparttable}
		\caption{\label{tab:table10}\footnotesize{The hfs constants $A$, $B$ of the ${^3\!}D_{1}$ and ${^3\!}F_{2}$ ionic states, and $A$ of the ${^2\!}D_{3/2}$ atomic level, given in MHz.}}
		\setlength{\tabcolsep}{0.7mm}{
			\begin{tabular}{cllll|l}\hline\hline
				\multicolumn{5}{c}{COLLAPS}  & \multicolumn{1}{c}{CRIS}\\\hline
				$A$&$A({^3\!}D_{1})$&$A({^3\!}F_{2})$&$B({^3\!}D_{1})$&$B({^3\!}F_{2})$& $A({^2\!}D_{3/2})$\\\hline
				45&$-$480.0(2)&368.5(1)&$-$11.4(6)&$-$54.9(9)&269.4(4)\\
				&$-$479.9(5)$^{\rm a}$&368.3(3)$^{\rm a}$&$-$12.6(19)$^{\rm a}$&$-$61.7(32)$^{\rm a}$&269.56(2)$^{\rm b}$\\
				46&$-$268.2(2)&205.7(1)&$+$8.0(6)&$+$30.8(10)&150.5(2)\\
				& & & & &150.576(9)$^{\rm c}$\\
				47&$-$526.2(2)&403.7(1)&$-$10.2(7)&$-$49.8(11)& \\
				49&$-$559.2(5)&429.3(3)&$-$8.2(26)&$-$39.7(19)&314.0(4)\\
				\hline\hline
		\end{tabular}}
		\begin{tablenotes}
			\item[a] These values for $^{45}$Sc isotope are taken from Ref.~\cite{Avgoulea_2011}.
			\item[b] This value for $^{45}$Sc isotope is taken from Ref.~\cite{45Sc-AB}.
			\item[c] This value for $^{46}$Sc isotope is taken from Ref.~\cite{46Sc-AB}.
		\end{tablenotes}
	\end{threeparttable}
\end{table}

\begin{table*}
	\begin{threeparttable}
		\centering \caption{The electromagnetic moments of Sc isotopes measured in this work using COLLAPS (in bold) and CRIS (in bold and italic) methods, compared to the literature values. The consistent CLS results are the recommended values, as discussed in the text.}
		\label{tab:table11}
		\setlength{\tabcolsep}{4.3mm}{
			\begin{tabular}{cccllll}\hline\hline
				&&&NMR or ABMR &CLS&ABMR&CLS\\\hline
				$A$&$I^{\pi}$& $T_{1/2}$&\multicolumn{2}{c}{$\mu$ ($\mu_{\rm N}$)}&\multicolumn{2}{c}{$Q_{\rm s,exp}$ (b)}\\
				\hline
				41&7/2$^-$& 596.3(17)~ms&+5.4283(14)~\cite{u-41Sc}$^{\rm a}$& &$-$0.145(3)~\cite{Q-41Sc}$^{\rm b}$ & \\
				43&7/2$^-$& 3.891(12)~h&+4.61(4)~\cite{47Sc} & $+$4.526(10)~\cite{Avgoulea_2011}$^{\rm a}$ & $-$0.26(6)~\cite{47Sc}$^{\rm b}$ & $-$0.27(5)~\cite{Avgoulea_2011} \\
				
				45&7/2$^-$&Stable&+4.75400(8)~\cite{45Sc-CPL}& &$-$0.220(2)~\cite{Q-45Sc}& \\
				
				46&4$^+$&83.79(4)~d&+3.03(2)~\cite{46Sc-AB}& +3.040(8)~\cite{Avgoulea_2011}$^{\rm a}$ & $+$0.119(2)~\cite{46Sc-AB}$^{\rm b}$& +0.12(2)~\cite{Avgoulea_2011}\\
				
				&     &     &      & \bf{+3.035(2)}/\bf{\textit{+3.036(6)}} & & \bf{+0.124(5)} \\
				47&7/2$^-$&3.3492(6)~d&+5.34(2)~\cite{47Sc}$^{\rm a}$&\textbf{+5.209(2)}&$-$0.22(4)~\cite{47Sc}$^{\rm b}$&\textbf{$-$0.200(6)} \\
				
				49&7/2$^-$& 57.18(13)~m&(+)5.61(3)~\cite{49Sc}$^{\rm a}$&\bf{+5.539(4)}/\bf{\textit{+5.540(11)}}& &\bf{$-$0.159(8)} \\
				
				\hline\hline
		\end{tabular}}
		\begin{tablenotes}
			\item[a] These are the recommended magnetic moments from Ref.\cite{Stone2019}.
			\item[b] These values are all re-evaluated relative to $Q$($^{45}$Sc) = $-$0.220(2) from Ref.~\cite{Q-45Sc}.
		\end{tablenotes}
	\end{threeparttable}
\end{table*}

\begin{figure*}[!t]
	\includegraphics[width=1.0\textwidth]{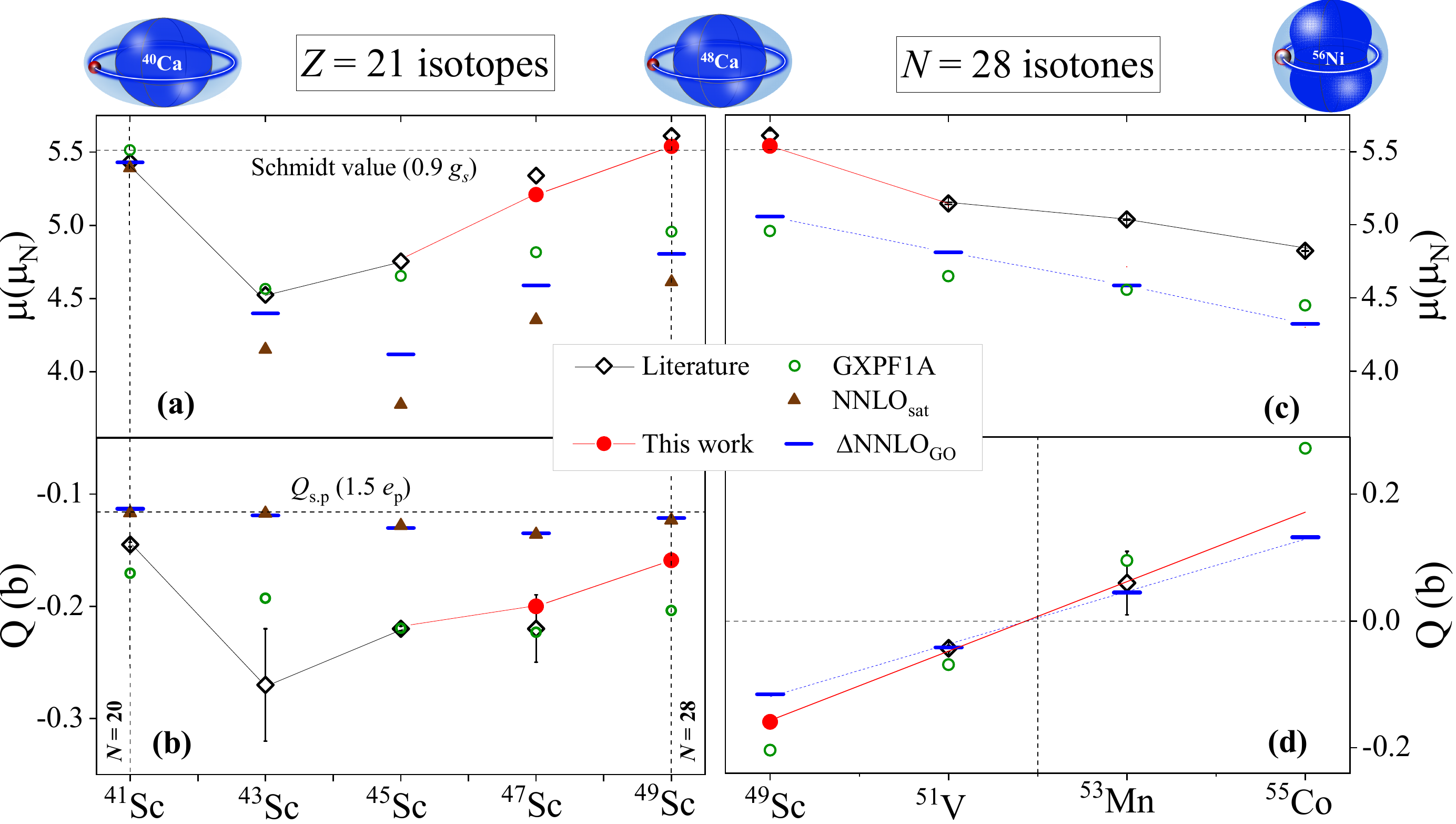}
	\caption{\label{fig:fig2}\footnotesize{(a) Magnetic and (b) quadrupole moments of $^{41-49}$Sc compared to SM GXPF1A calculations~\cite{GXPF1A} using the effective $g$-factor ($g_s^{\rm eff}=0.9g_s^{\rm free}$) and effective charges ($e_{\rm \pi}=1.5e$, $e_{\rm \nu}=0.5e$) and to VS-IMSRG calculations (using free $g$-factors and bare nucleon charges) based on the NNLO$_\mathrm{sat}$~\cite{NNLOsat} and $\Delta$NNLO$_\mathrm{GO}$~\cite{NNLOgo} interactions. Dashed horizontal lines indicate the effective SP magnetic and quadrupole moments of one proton in the $\pi f_{7/2}$ orbit. (c) Magnetic and (d) quadrupole moments of $N=28$ isotones, compared to theoretical calculations. $^{51}$V, $^{53}$Mn and $^{55}$Co moments are taken from Refs.~\cite{u-51V,u-53Mn,u-55Co,Q-51V,Q-53Mn}. The red line in (d) is a linear fit of the data. Three sketches on the top with exaggerated deformation represent the shapes of $^{41,49}$Sc with a proton outside $^{40,48}$Ca and of $^{55}$Co with a hole inside $^{56}$Ni.}}
\end{figure*}

Table~\ref{tab:table11} presents the newly-measured electromagnetic moments of Sc isotopes, along with the literature values~\cite{Avgoulea_2011,47Sc,45Sc-CPL,Q-45Sc,u-41Sc,Q-41Sc,49Sc}. The new results of $^{47,49}$Sc are more precise than those from atomic-beam magnetic resonance (ABMR) and nuclear magnetic resonance (NMR) experiments~\cite{47Sc,49Sc}. A systematic deviation of $\sim$2$\%$ is found between the magnetic moments of $^{43,47,49}$Sc measured using CLS and those measured using NMR~\cite{49Sc} or ABMR~\cite{47Sc}. The magnetic moments of $^{43,47}$Sc were measured in one ABMR experiment~\cite{47Sc}, and the $^{49}$Sc magnetic moment was obtained via NMR and determined relative to the $^{47}$Sc ABMR moment~\cite{49Sc}, which links all these moments. The magnetic moments measured with CLS were obtained from three independent experiments at IGISOL~\cite{Avgoulea_2011}, COLLAPS and CRIS (this work), using four ionic and one atomic states, and are all in excellent agreement. The $^{46}$Sc magnetic moment was, however, measured in another independent ABMR experiment~\cite{46Sc-AB}, which is in excellent agreement with the CLS results. This indicates that the discrepancy between the newly measured $^{43,47,49}$Sc magnetic moments and those from ABMR/NMR methods can all be traced back to one particular ABMR experiment~\cite{47Sc}.

\section{Discussion}
\label{sec4}

Our measurements provide key experimental data for the systematical investigation of the magnetic and quadrupole moments of $Z=21$ isotopes and $N=28$ isotones when valence nucleons fill the unique $f_{7/2}$ orbit, as presented in Fig.~\ref{fig:fig2}. Particularly, the first $Q_{\rm s}$ measurement of the short-lived $^{49}$Sc is essential to validate the simple seniority scheme of the independent-particle SM (see Fig.~\ref{fig:fig2}(d) and discussion below).

As presented in Fig.~\ref{fig:fig2}(a, b), both magnetic and quadrupole moments of $^{41-49}$Sc \mbox{($Z=21$, even $N=20-28$)}, characterized with the identical parabolic trends, approach the SP values for a proton in the $\pi f_{7/2}$ orbit at the neutron magic numbers 20 and 28. This points to the rather pure SP character of $^{41,49}$Sc and the doubly-magic nature of $^{40,48}$Ca. Note that the SP moments are calculated using an effective $g_s$-factor of $g_s^{\rm eff}$ = $0.9g_s^{\rm free}$ and an effective charge of $e_{\rm \pi}=1.5e$, to compensate for the possible missing core excitations. The deviation of the magnetic moment of $^{43-47}$Sc~\mbox{($N=22, 24, 26$)} from the Schmidt line indicates an enhancement of NN correlation as more neutrons/holes are added to the $f_{7/2}$ orbit. The single proton outside the $^{40}$Ca and $^{48}$Ca cores induces an oblate core polarization for $^{41,49}$Sc (sketches on the top of Fig.~\ref{fig:fig2}) leading to a negative $Q_{\rm s}$ (Fig.~\ref{fig:fig2}(b)). This core polarization effect is maximized around mid-shell where more particles/holes appear in the $\nu f_{7/2}$ orbit, but a more precise measurement of the $^{43}$Sc quadrupole moment should confirm the expected quadratic trend for these quadrupole moments.

For the $N=28$ isotonic sequence with (odd-) protons filling the $\pi f_{7/2}$ orbit from Sc ($Z=21$) to Co ($Z=27$), the magnetic moment is expected to be constant (\lq Schmidt value' in Fig.~\ref{fig:fig2}(c)) from the independent-particle SM. However, the experimental values follow a characteristic linear deviation from the SP value. This phenomenon can be explained as due to increasing cross-shell proton excitations to the upper $f_{5/2}$ spin-orbit partner when the $\pi f_{7/2}$ orbit is being filled~\cite{Arima1954,Yoshina1972}. A minor mixing of this $M1$-excitation into the odd-proton wave function may have a large impact on the magnetic moment~\cite{Yoshina1972,TOWNER1987}. Thus, the magnetic moment ($\mu$) of an isotone with $n$ protons ($\pi f^n_{7/2}$) follows a linear trend proportional to $n$ and a constant $\delta \mu$ that relates to the $M1$ spin-flip matrix element: \mbox{$\mu(\pi f_{7/2}^n)=\mu(^{49}{\rm Sc})+(n-1)\delta\mu$}. As a result, a fraction of such orbit mixing in $^{55}$Co induces the observed reduction of its $\mu$ relative to $\mu(^{49}{\rm Sc})$, further emphasizing the relatively \lq pure' SP nature of $^{49}$Sc.

In the extreme independent-particle SM, the seniority scheme allows an estimation of the spectroscopic quadrupole moment of isotones with odd protons ($n$) filling an orbit $j$~\cite{Heyde1994,Castel1990}:
\begin{equation}
	Q_{\rm{s}}=\left(1-\frac{2n-2}{2j-1}\right) Q_{\rm{s.p}}(j)
	\label{eq:one}
\end{equation}
leading to a linear increase proportional to $Q_{\rm{s.p.}}(j)$, the SP quadrupole moment for a proton in the orbit $j$~\cite{Heyde1994,Castel1990,RevModPhys-moments}. This linear trend of the $Q_{\rm{s}}$ is expected to cross zero when the orbit $j$ is half filled. The single proton outside the doubly-magic $^{48}$Ca core induces an oblate core polarization for $^{49}$Sc (negative $Q_{\rm s}$), whereas a prolate shape is predicted with a positive $Q_{\rm s}$ for $^{55}$Co~(due to a hole inside the doubly-magic $^{56}$Ni), as schematically presented on the top of Fig.~\ref{fig:fig2}. With the addition of the present precise measurement of $Q_{\rm s}$ for $^{49}$Sc, a linear trend can then be unambiguously determined from the available experimental $Q_{\rm{s}}$ of $N=28$ isotones, crossing zero at the half-filling of the $\pi f_{7/2}$ orbit~(Fig.~\ref{fig:fig2}(d)), representing a textbook example for the independent-particle SM picture. It is worth noting that the proton cross-shell excitations, which strongly affect the magnetic moments~($M1$ correlations) (Fig.~\ref{fig:fig2}(c)), have no notable effect on the quadrupole moments~($E2$ correlations).

Naturally, one would expect the large-scale SM to give a good description of the above discussed nuclear moments. However, as shown in Fig.~\ref{fig:fig2}(a, c), the SM calculation using the GXPF1A effective interaction ($^{40}$Ca core and $pf$ model space)~\cite{GXPF1A}, does not reproduce well the trend of the magnetic moments for heavier scandium isotopes, and systematically underestimates those of the $N=28$ isotones. This may suggest a missing polarization effect of the $^{40}$Ca core, requiring an effective interaction with a valence space that includes $sd$ and $pf$ shells and is optimized for the calcium region. While interactions exist for protons and neutrons in the $sdpf$ model space (e.g. SDPF-U~\cite{SDPF-U} and SDPF-MU~\cite{SDPF-MU}), these interactions have been developed for isotopes with $Z=8-20$ (namely in the $sd$-shell) to properly account for the neutron $sd-pf$ shell excitations in their neutron-rich isotopes. Thus, an interaction that properly takes into account proton excitations across $Z=20$ yet needs to be developed.

\begin{figure}[t!]
	\includegraphics[width=0.49\textwidth]{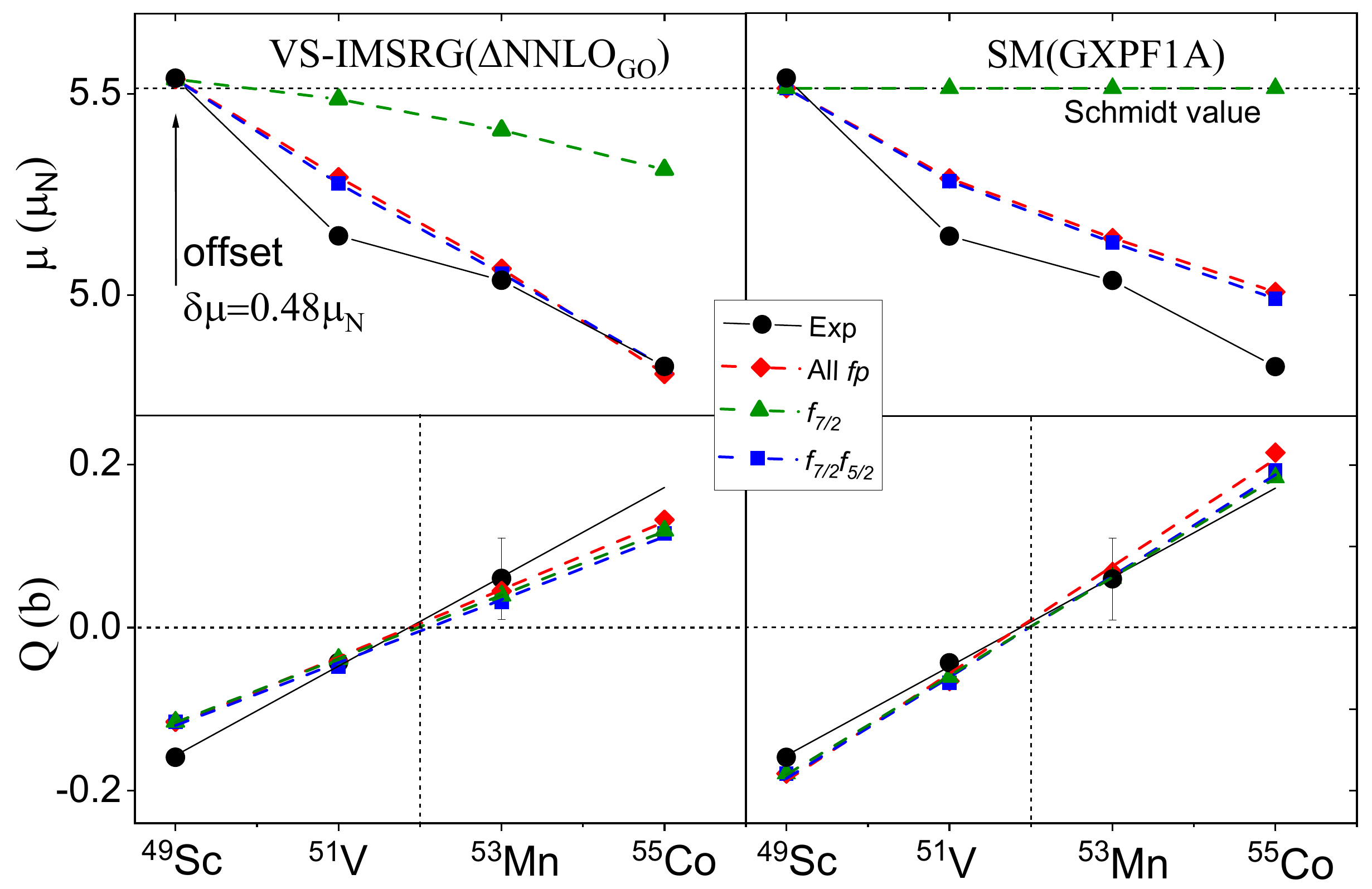}
	\caption{\label{fig:fig3}\footnotesize{Nuclear moments of $N=28$ isotones calculated with VS-IMSRG and SM with protons constrained on different orbits, and with neutron excitation forbidden. For a better visualization, an intended offset $\delta \mu =0.48$ is added to the VS-IMSRG values.}}
\end{figure}

With advances in many-body methods and NN+3N forces from $\chi$EFT~\cite{Machleidt2011,Epelbaum2009}, first-principles calculations of electromagnetic properties of medium-mass nuclei are now possible using the VS-IMSRG~\cite{IMSRG-Stroberg2017,Morris2015,Stro19ARNPS,Miya20MS}, with first applications in the $sd$ shell~\cite{37Ca-moment,Al-moment,Hend18E2,Heil19OE2,Ciem20OE2,Hend20E2,Garns17M3}. Here, we use two chiral interactions for the first time in the $pf$ space: NNLO$_\mathrm{sat}$~\cite{NNLOsat} and $\rm{\Delta NNLO_{GO}}$(394)~\cite{NNLOgo}; the latter includes explicitly $\rm{\Delta}$(1232)-isobars and has so far only been tested for charge radii and binding energies~\cite{Kradii,CCNNLOgo}. In this work, the VS-IMSRG calculation follows the same procedure as in Ref.~\cite{37Ca-moment}, with an increased $E_\mathrm{3max} = 22$ truncation on storage of 3N matrix elements~\cite{Miya21Heavy}. We decouple a $pf$-shell valence-space Hamiltonian above a $^{40}$Ca core (or $^{48}$Ca core for $N=28$ isotones), and the $E2$ and $M1$ operators are consistently transformed by the VS-IMSRG~\cite{Parz17Trans} to produce consistent effective valence-space operators. Final energies and transition rates are obtained with the KSHELL code~\cite{SHIMIZU2019372}. We emphasize that only bare nucleon charges and free $g$-factors are used here, which is fundamentally different from the SM GXPF1A calculation where $g_s^{\rm eff}=0.9g_s^{\rm free}$ and $e_{\rm \pi}=1.5e$, $e_{\rm \nu}=0.5e$ have been used.

Similar to the SM calculation, both chiral interactions ($\rm{\Delta NNLO_{GO}}$ and $\rm NNLO_{sat}$) result in a clear underestimation of the magnetic moments of $^{45,47,49}$Sc, and a systematic underestimation of magnetic moment trend of the $N=28$ isotones, as shown \mbox{(Fig.~\ref{fig:fig2}(a, c))}. We note that the new $\rm{\Delta NNLO_{GO}}$ interaction gives a somewhat better description along the Sc isotopic chain, which may benefit from the inclusion of the $\rm{\Delta}$(1232)-isobar degree of freedom.

In \mbox{Fig.~\ref{fig:fig2}(b, d)}, the quadrupole moments are compared with the calculated moments from these theories. While the SM GXPF1A calculations follow reasonably well the general trend of $Q_{\rm s}$ as a function of $N$, the VS-IMSRG calculations largely underestimate the absolute value of the experimental quadrupole moments of $Z=21$ isotopes \mbox{(Fig.~\ref{fig:fig2}(b))}. A similar underestimation was already seen for calculated $E2$ matrix elements in lighter Mg isotopes~\cite{Hend18E2,Hend20E2}, which is likely due to missing higher-order collective excitations in the VS-IMSRG calculation at the IMSRG(2) level, as discussed in Refs.~\cite{Hend18E2,Hend20E2}. Further theoretical studies are needed to understand the origin of missing $E2$ correlations in the \textit{ab-initio} calculations. For the $N=28$ isotones, as shown in \mbox{Fig.~\ref{fig:fig2}(d)}, the characteristic symmetric linear trend of the $Q_\textrm{s}$ when filling the proton $\pi f_{7/2}$ orbit is captured remarkably well by both theories. The smaller slope of the calculated trend using the \textit{ab-initio} interaction reflects only a small underestimation of the SP quadrupole moment for a proton (particle or hole) in the $f_{7/2}$ orbit (no effective charges are used here). With the GXPF1A effective interaction, the slope is overestimated, which points to either a too large effective proton charge, or an overestimated value for the SP quadrupole moment. This can be further investigated by blocking neutron excitations across $N=28$.

We further investigate the influence of proton $M1$ and $E2$ correlations in the $pf$ model space on the moments of $N=28$ isotones by performing calculations for protons in a gradually extended model space, as shown in Fig.~\ref{fig:fig3}. For consistency with the VS-IMSRG calculation, neutron excitations across $N=28$ are intentionally blocked in the SM calculations. As presented in Fig.~\ref{fig:fig3}~(upper panel), clearly, proton excitations to the $\pi f_{5/2}$ are driving the linearly increased deviation of $\mu$ from the Schmidt value from $^{49}$Sc to $^{55}$Co, and this is captured well in both theories. On the contrary, these proton $M1$-excitations have nearly negligible impact on the $Q_{\rm s}$ moments (Fig.~\ref{fig:fig3}~(lower panel)). The SM GXPF1A calculations for magnetic and quadrupole moments without neutron excitations across $N = 28$ (Fig.~\ref{fig:fig3}, right panel), show a better agreement with the experimental data, in comparison to the SM results performed in the full model space \mbox{(Fig.~\ref{fig:fig2}(c, d))}. This suggests that neutron excitations across $N=28$, which in turn correlatively induce proton excitations across $Z=28$, are overestimated in the full SM calculations~(Fig.~\ref{fig:fig2}) and a small portion of these excitations leads to a notable change of $\mu$ as discussed above, supporting the \lq pure' SP character of $^{49}$Sc. In other words, the $N=28$ shell gap seems to be underestimated in the GXPF1A shell model interaction.

It would be worth noting that there is a substantial difference between the performance of the theories for the $N=28$ isotones~(Fig~\ref{fig:fig2}(c, d) and Fig.~\ref{fig:fig3}) and that for the $Z=21$ isotopes (Fig.~\ref{fig:fig2}(a, b)). Both SM and VS-IMSRG calculations give a good description of the systematic trends of moments for $N=28$ isotopes when odd protons fill the $f_{7/2}$ orbit~(Fig~\ref{fig:fig2}(c, d) and Fig.~\ref{fig:fig3}). This benchmarks the significant progress of the \textit{ab initio} calculations for the description of electromagnetic moments of the simple cases with magic neutron number ($N=28$) and valences protons outside the doubly-magic $^{48}$Ca core. As for $^{41-49}$Sc isotopes where enhanced NN correlations are induced by the additional neutron/holes in the $f_{7/2}$ orbit and the possible polarization effect of $^{40}$Ca core, the theories are much less successful in describing their electromagnetic moments. This, instead, provides a systematical quantification of their deviations from experimental data along the entire isotopic chain, motivating further development of the nuclear interactions and the many-body methods when encountering more complicated correlations.

\section{Summary and conclusion}
In summary, electromagnetic moments of $^{47,49}$Sc were measured with improved precision and accuracy, among with the electric quadrupole moment $Q_{\rm s}$ of $^{49}$Sc isotope obtained for the first time. A systematic investigation of electromagnetic moments has been performed for $N=28$ isotones and $Z=21$ isotopes with valence nucleons filling the $f_{7/2}$ orbit. Thanks to the unique location of this orbit in the SM scheme, the sensitivity of electromagnetic moments to the nucleon-nucleon $M1$ and $E2$ correlations is probed. Particularly the seniority scheme of the independent-particle SM is experimentally confirmed based on the $Q_{\rm s}$ of $^{49}$Sc and its $N = 28$ isotones, providing a textbook example. This study serves as a benchmark for state-of-the-art theoretical models, especially \textit{ab initio} VS-IMSRG calculations using microscopic interactions derived from $\chi$EFT. At the level of the experimental precision, none of the theories used in this work satisfactorily reproduces the magnetic moment trends along the $Z=21$ isotopic chain or their absolute values for the $N=28$ isotones. In particular towards $N=28$, all magnetic moments are largely underestimated, which may suggest that neutron $M1$ excitations to the $f_{5/2}$ are too pronounced in the models. As for the quadrupole moments, the trend along the $Z=21$ isotopic chain is reasonably reproduced by the SM GXPF1A calculations, although the absolute value towards $N=28$ is clearly overestimated, suggesting that $E2$ excitations to the $p_{3/2}$ orbit are also overestimated. Together with the underestimated magnetic moments, this points to a too small $N=28$ gap in the GXPF1A interaction. The linear trend of $Q_{\rm s}$ observed for $N=28$ isotones is well described with the SM GXPF1A, but the absolute SP quadrupole moment (or the effective charge) is overestimated, as the slope is too steep. With the \textit{ab-initio} interactions, the quadrupole moments of $Z=21$ isotopes are largely underestimated, pointing to missing $E2$ correlations when opening the neutron shell between $N = 20$ and 28. Nevertheless, the linear trend of quadrupole moments along $N=28$ isotones is very well captured, and the slope is only a little less steep than observed, illustrating that the SP quadrupole moment (without use of effective charge) is well reproduced by the \textit{ab-initio} theory. The present work highlights the progress made in advanced nuclear theory. It paves the way for a coherent description of basic nuclear properties with further development of the nuclear interactions and the many-body methods, e.g. a more proper effective SM interaction for the calcium region in the $sdpf$ model space, VS-IMSRG approach with all operators truncated at the three-body level and decoupling a cross-shell Hamiltonian as well as inclusion of MEC.

\section*{Acknowledgments}

We acknowledge the support of the ISOLDE collaboration and technical teams and S. R. Stroberg for the imsrg++ code~\cite{Stro17imsrg++} used to perform VS-IMSRG calculations. This work was supported by the National Key R\&D Program of China (Contract No. 2018YFA0404403), the National Natural Science Foundation of China (No:11875073, U1967201, 11775316); the BriX Research Program No. P7/12, FWO-Vlaanderen (Belgium), GOA 15/010 from KU Leuven; the UK Science and Technology Facilities Council grants ST/L005794/1 and ST/P004598/1; ERC Consolidator Grant No.648381 (FNPMLS); the NSF grant PHY-1068217, the BMBF Contract No. 05P18RDCIA; the Max-Planck Society, the Helmholtz International Center for FAIR (HIC for FAIR); the EU Horizon2020 research and innovation programme through ENSAR2 (no. 654002), NSERC under grants SAPIN-2018-00027 and RGPAS-2018-522453 and the Arthur B. McDonald Canadian Astroparticle Physics Research Institute. TRIUMF receives funding via a contribution through the National Research Council of Canada and computations of VS-IMSRG were performed with an allocation of computing resources on the Cedar at WestGrid and Compute Canada.

\bibliographystyle{elsarticle-num}
\bibliography{ref-Sc}

\end{document}